\documentclass[twocolumn,showpacs,superscriptaddress,preprintnumbers,amsmath,amssymb]{revtex4}

\usepackage{graphicx}
\usepackage{dcolumn}
\usepackage{bm}

\newcommand{\pstar}{\mathrm{P}^*}
\newcommand{\ud}{\mathrm{d}}
\newcommand{\ui}{\mathrm{i}}
\newcommand{\ue}{\mathrm{e}}
\newcommand{\es}{E^*}
\newcommand{\VaR}{\Lambda^*}

\begin{document}

\preprint{FNT/T 2009/04}

\title{A Generalized Fourier Transform Approach to Risk Measures}

\author{Giacomo Bormetti}
\email{Giacomo.Bormetti@pv.infn.it}
\affiliation{
  Istituto Universitario di Studi Superiori, Centro Studi Rischio e Sicurezza\\
  Viale Lungo Ticino Sforza 56, 27100 Pavia, Italy
}
\affiliation{
  Istituto Nazionale di Fisica Nucleare, Sezione di Pavia\\
  Via Bassi 6, 27100 Pavia, Italy
}
\author{Valentina Cazzola}
\affiliation{
  Istituto Universitario di Studi Superiori, Centro Studi Rischio e Sicurezza\\
  Viale Lungo Ticino Sforza 56, 27100 Pavia, Italy
}
\affiliation{
  Dipartimento di Fisica Nucleare e Teorica, Universit\`a degli Studi di Pavia\\ 
  Via Bassi 6, 27100 Pavia, Italy
}
\affiliation{
  Istituto Nazionale di Fisica Nucleare, Sezione di Pavia\\
  Via Bassi 6, 27100 Pavia, Italy
}
\author{Giacomo Livan}
\affiliation{
  Dipartimento di Fisica Nucleare e Teorica, Universit\`a degli Studi di Pavia\\ 
  Via Bassi 6, 27100 Pavia, Italy
}
\affiliation{
  Istituto Nazionale di Fisica Nucleare, Sezione di Pavia\\
  Via Bassi 6, 27100 Pavia, Italy
}
\author{Guido Montagna}
\affiliation{
  Dipartimento di Fisica Nucleare e Teorica, Universit\`a degli Studi di Pavia\\ 
  Via Bassi 6, 27100 Pavia, Italy
}
\affiliation{
  Istituto Nazionale di Fisica Nucleare, Sezione di Pavia\\
  Via Bassi 6, 27100 Pavia, Italy
}
\affiliation{
  Istituto Universitario di Studi Superiori, Centro Studi Rischio e Sicurezza\\
  Viale Lungo Ticino Sforza 56, 27100 Pavia, Italy
}
\author{Oreste Nicrosini}
\affiliation{
  Istituto Nazionale di Fisica Nucleare, Sezione di Pavia\\
  Via Bassi 6, 27100 Pavia, Italy
}
\affiliation{
  Istituto Universitario di Studi Superiori, Centro Studi Rischio e Sicurezza\\
  Viale Lungo Ticino Sforza 56, 27100 Pavia, Italy
}

\date{\today}

\begin{abstract}
We introduce the formalism of generalized Fourier transforms in the context of risk management. We develop a general framework to efficiently compute 
the most popular risk measures, Value-at-Risk and Expected Shortfall (also known as Conditional Value-at-Risk). The only ingredient required by our approach
is the knowledge of the characteristic function describing the financial data in use. This allows to extend risk analysis to those non-Gaussian models defined in 
the Fourier space, such as L\'evy noise driven processes and stochastic volatility models.  
We test our analytical results on data sets coming from various financial indexes, finding that our predictions outperform those provided by the 
standard Log-Normal dynamics and are in remarkable agreement with those of the benchmark historical approach. 
\end{abstract}


\maketitle

\section{Introduction}

September 2008 financial crisis has dramatically highlighted the need for reliable, easy to understand and implement instruments to measure and manage risk.
The high volatility of financial markets during the Nineties induced academics and practitioners to design sophisticated risk management tools.
According to the recently revised capital adequacy framework, commonly known as Basel~II accord \cite{Basel:2006}, any financial institution has to meet stringent
capital requirements in order to cover the various sources of risk to be faced as a result of normal operations. 
Today the most widely used measure to manage market risk in the financial industry is Value-at-Risk (VaR).
VaR refers to the maximum potential loss over a given period at a certain
confidence level (CL) and can be used to measure the risk of individual assets and portfolios of assets as well.
Because of its conceptual simplicity, VaR has been extensively adopted by regulators and it generally provides a reasonably accurate estimate of risk.
However, VaR is known to suffer from important drawbacks: it can violate the sub-additivity rule for portfolio risk, which is a required
property for any consistent measure of risk \cite{Jorion:2001,Acerbi_Tasche:2001}, and it does not quantify the typical loss 
incurred when the risk threshold is exceeded.
The expected shortfall (ES), defined as the expected loss conditional on the VaR threshold being exceeded, overcomes 
these disadvantages and leads to more consistent results. \\
\indent Three main approaches are known in the literature and used in practice to compute VaR and ES: the parametric approach, the historical one, and Monte Carlo simulations of the stochastic dynamics of a given stock price returns model.
The parametric approach usually relies on the normality assumption for the returns distribution, although some analytical
results using non-Gaussian functional forms are available in the literature \cite{Jorion:2001,Heikkinen_Kanto:2002,Kamdem:2005,Bormetti_etal:2007}. However, it is well known
that empirical price returns exhibit heavy tails and a certain degree of asymmetry; 
the historical simulation approach is often used in order to capture their leptokurtic nature. 
The last approach consists of Monte Carlo simulations of the return dynamics, but it usually requires very intensive simulations to get to acceptably accurate risk estimates. As a result of the present situation, reliable and hopefully fast methods to calculate financial risk are mandatory. \\
\indent In this article we shall present a general framework to compute VaR and ES by only relying on the knowledge of the closed-form characteristic function (CF) 
describing the distribution of the financial returns under analysis. Our approach draws on the original ideas developed in \cite{Lewis:2001,Lipton:2001} by A. L. Lewis and A. Lipton, who introduced
generalized Fourier calculus in the context of option pricing, and extends them to the risk management framework.
The advantage of our approach is manifold. The equations we obtain for risk measures are intuitive and easy to read, since both VaR and ES turn out to be expressed in terms of one same function evaluated for different arguments. Moreover, the evaluation of such formulae is computationally efficient, since running twice a 
fast Fourier transform (FFT) algorithm yields both VaR and ES values over wide ranges of the CL. This is a remarkable feature, since under standard 
approaches risk measures would need to be recomputed every time the CL is changed. Fourier inversion based approaches can be found in literature; 
the first attempt dates back to the work of Rouvinez \cite{Rouvinez:1997}, and later developments are discussed in 
\cite{Cardenas_etal:1998,Britten-Jones_Schaefer:1999,Mina_Ulmer:1999,Duffie_Pan:2001,Glasserman_etal:2002,Albanese_etal:2004}. 
However, the Fourier inversion is usually employed to compute an approximation of the cumulative function through the Inversion Theorem \cite{Gil-Pelaez:1951},
then the quantile corresponding to the fixed CL is computed by root-finding algorithms. However, this final step has to be iterated over the entire set of desired CLs,
while our approach directly provides the risk estimates.
A very useful and easy to interpret graphical representation of the results can also be sketched. 
Finally, being based on the use of CFs, our method is readily applicable to a number of interesting 
distributions whose probability density function (PDF) is not known analytically. 
Remarkable examples are represented by the class of L\'evy distributions, both in their original and truncated versions.

In this article we focus on Truncated L\'evy Distributions (TLDs), which, given their ability to reproduce some of the stylized facts observed in real market data, 
have been introduced and applied in the context of financial analysis by several authors \cite{Mantegna_Stanley:1994,Koponen:1995,Matacz:2000,Mantegna_Stanley:2000}. 
The approach we propose can also deal with stochastic volatility models (SVMs), which have already been successfully used in the context of derivative pricing. 
Interestingly, to the best of our knowledge, these models have never been employed in risk management before. Our framework is naturally suited to models
that are well defined in terms of the CF such as the Stein-Stein \cite{Stein_Stein:1991,Masoliver_Perello:2002}, 
Heston \cite{Heston:1993,Dragulescu_Yakovenko:2002}, Sch\"obel-Zhu \cite{Schobel_Zhu:1999}, and
exponential Ornstein-Uhlenbeck \cite{Scott:1987,Masoliver_Perello:2006,Bormetti_etal:2008,Bormetti_Cazzola_Delpini:2009} models.
We choose to work with the Heston model, whose popularity is rapidly growing amongst financial practitioners \cite{Bormetti_Livan:2009}. 
A further feature of SVMs we wish to investigate in the present work is their ability to provide high order normalized cumulants with different time scalings w.r.t. 
those implied by the Central Limit Theorem (CLT). We plan to test their performances when projecting risk estimates over time.

The paper is organized as follows: in Section II the technical definitions of VaR and ES are recalled, and their expressions in terms of the CF are derived. 
In Section III the models we use to test our approach are presented. In Section IV the fitting procedures and the data analysis
we performed are described, and the numerical results obtained for the risk measures are detailed. In Section V some conclusions are drawn.

\section{Formulae for risk estimation}

The Value-at-Risk (VaR) is defined as the maximum potential loss $-\Delta^*$ (with $\Delta^* > 0$) not to be exceeded at 
a given significance level $\pstar \in (0,1)$ over a fixed time horizon $\Delta t$. Thus, when considering the price variation 
$\Delta S = S - S_0$ of some asset or index, VaR is implicitly defined by the integral equation
\begin{equation} \label{VaR1}
  \pstar = \int_{-S_0}^{-\Delta^*} \ud (\Delta S) \ \hat{p}(\Delta S)
\end{equation}
where $\hat{p}$ is the PDF describing $\Delta S$. 
When switching to the centered logarithmic returns $x \doteq \ln(1+\Delta S/S_0 ) - \mu \Delta t$, with $\mu$ the 
linear returns mean, equation (\ref{VaR1}) can be rewritten as
\begin{equation} \label{VaR2}
  \pstar = \int_{-\infty}^{-L^*} \ud x \ p(x) 
\end{equation}
where $p$ is the PDF associated with $x$ and $L^* \doteq - \ln (1 - \Delta^*/S_0) + \mu \Delta t$. 
It is worth mentioning that $\Delta^*$ represents VaR in monetary units; we shall often use the normalized VaR $\Lambda^* \doteq \Delta^* / S_0$, 
which is usually presented as a percentage quantity (percentage VaR).

In order to derive new expressions for VaR, we adapt the approach developed by A. L. Lewis and, independently, by A. Lipton
in the context of derivative pricing under stochastic volatility \cite{Lewis:2001,Lipton:2001}. 
Let us represent the PDF $p$ in terms of the generalized Fourier transform $f$
\begin{equation*}
  p(x) = \frac{1}{2\pi} \int_{-\infty + \ui \nu}^{+\infty + \ui \nu} \ud \phi \ f(\phi) \ \ue^{-\ui \phi x}.
\end{equation*}
In this integral expression $\phi = \omega + \ui \nu$ is a complex variable ($\omega$, $\nu \in \mathbb{R}$), 
whose imaginary part $\nu$ belongs to the proper strip of regularity $(\nu_-,\nu_+)$ of the extended characteristic function (ECF) $f$. 
Such a strip is delimited by the possible singularities of $f$ (which can be shown to be purely imaginary under suitable conditions \cite{Lewis:2001}) 
lying closest to the origin in the complex upper (lower) half plane, whose imaginary part reads $\nu_+$ ($\nu_-$). 
With these positions, we can plug the previous expression in equation ($\ref{VaR2}$) and switch the integration order. Thus, we obtain
\begin{equation*}
  \pstar = \frac{1}{2\pi}\int_{-\infty + \ui \nu}^{+\infty + \ui \nu} \ud \phi \ f(\phi) \int_{-\infty}^{-L^*} \ud x \ \ue^{-\ui \phi x}
\end{equation*}
and to ensure convergence of the second integral we require $\nu$ to be strictly positive. 
So, restricting $\nu \in (0,\nu_+)$ equation (\ref{VaR2}) eventually becomes
\begin{eqnarray}
  \pstar &=& \frac{\ui}{2\pi}  \int_{-\infty + \ui \nu}^{+\infty + \ui \nu} \ud \phi 
  \ f(\phi) \frac{\ue^{\ui \phi L^*}}{\phi} \nonumber \\ &=& \frac{\ue^{-\nu L^*}}{2\pi} 
  \int_{-\infty}^{+\infty} \ud \omega \ \frac{f(\omega + \ui \nu) \ \ue^{\ui \omega L^*}}{\nu - \ui \omega}\nonumber \\ 
  &=& \frac{\ue^{-\nu L^*}}{\pi} \mathrm{Re} \left [ \int_0^{+\infty} \ud \omega \ \frac{f(\omega + \ui \nu) 
  \ \ue^{\ui \omega L^*}}{\nu - \ui \omega} \right ]\nonumber
\end{eqnarray}
where the final equality is obtained by exploiting the symmetries of the real and imaginary parts of the ECF. 
Then, defining the function
\begin{equation} \label{G}
  G_{\nu} (L^*,\theta) \doteq \ue^{-\theta L^*} \int_0^{+\infty} \ud \omega \ \frac{f(\omega + \ui \nu)}{\theta - \ui \omega} \ue^{\ui \omega L^*}
\end{equation}
allows us to write
\begin{equation} \label{FFTVaR}
  \pstar = \frac{\mathrm{Re} \ G_{\nu}(L^*,\nu)}{\pi}.
\end{equation}
We now follow a similar line of reasoning for the Expected Shortfall (ES), defined as the average potential loss when the VaR threshold 
for a fixed $\pstar$ is exceeded.  With the same notation as above, in terms of the linear and centered logarithmic returns, we can write for the ES $E^*$
\begin{eqnarray}\label{eq:ES}
  \es(\pstar) &=& - \frac{1}{\pstar} \int_{-S_0}^{-\Delta^*(\pstar)} \ud (\Delta S) \ \Delta S \ \hat{p}(\Delta S) \nonumber \\
  &=& - \frac{S_0}{\pstar} \int_{-\infty}^{-L^*(\pstar)} \ud x \ p(x) \left (\ue^{x + \mu \Delta t} - 1 \right ),
\end{eqnarray}
where we have made explicit the dependence of $\es$, $\Delta^*$ and $L^*$ on $\pstar$. 
From now on we shall drop this dependence and we shall assume $S_0 = 1$. 
As we did before, we can plug the generalized Fourier transform of $p$ into the previous equation and switch the integration order
\begin{eqnarray} \label{ES2}
  \es &=& - \frac{\ue^{\mu \Delta t}}{2\pi\pstar} \int_{-\infty + \ui \nu}^{+\infty + \ui \nu} \ud \phi \ f(\phi) \nonumber \\ &\times&
  \int_{-\infty}^{-L^*} \ud x \left ( \ue^{(1-\ui \phi)x} - \ue^{-\mu \Delta t - \ui \phi x} \right ).
\end{eqnarray}
When we pose $\phi = \omega + \ui \nu$, the first of the two integrands in $\ud x$ requires $\nu > -1$ to be evaluated, while the second one requires $\nu > 0$. So, all in all, we are again left with $\nu \in (0,\nu_+)$ and, recalling the definition ($\ref{G}$) of $G_{\nu}$, equation ($\ref{ES2}$) reads
\begin{equation} \label{FFTES}
  \es = 1 - \ue^{\mu \Delta t} \frac{\mathrm{Re} \ G_{\nu}(L^*,\nu + 1)}{\mathrm{Re} \ G_{\nu}(L^*,\nu)}.
\end{equation}
Formulae ($\ref{FFTVaR}$) and (\ref{FFTES}) do represent the first main contribution of
this work and let us now see what the main advantage in their use is. 
Usually, VaR and ES are evaluated in correspondence of a \emph{single} fixed significance level $\pstar$ by means of different techniques, see 
\cite{Jorion:2001} and the on-line repository at \verb|www.gloriamundi.org| for an exhaustive review. 
In the best case scenario, the financial practitioner is provided with a closed-form expression dependent on distributional assumptions and 
parametric in the quantile of a standardized PDF and in few free parameters to be calibrated on the financial time series in use, see e.g. 
\cite{Mina_Xiao:2001,Bormetti_etal:2007,Bormetti_etal:2009}.   
However, every time the value of $\pstar$ is changed, risk measures need to be re-computed. Moreover, as anticipated in the introduction, up to now there are no
efficient ways to conjugate risk estimation with models fully characterized in terms of the CF. Indeed, this is the case for the class of L\'evy
stable distributions and their exponentially damped version \cite{Mantegna_Stanley:1994,Koponen:1995} and for all those dynamical models emerging 
in the context of option pricing under stochastic volatility,
such as the Stein-Stein, Heston, Schobel-Zhu, exponential Ornstein-Uhlenbeck models 
\cite{Scott:1987,Stein_Stein:1991,Heston:1993,Schobel_Zhu:1999,Masoliver_Perello:2002,Dragulescu_Yakovenko:2002,Masoliver_Perello:2006,Bormetti_etal:2008} 
and their extensions dealing with jump diffusion, e.g. \cite{Lipton_Sepp:2008}.
The use of the $G_{\nu}$ function can overcome all these drawbacks. As a matter of fact, once the ECF of the financial dynamics at hand is known in closed-form, 
a grid of $\omega$ values can be set and $G_{\nu}(L^*,\nu)$ (for an admissible value of $\nu$) can be efficiently evaluated via FFT algorithms. 
This leaves us with a vector $\bm{L^*}$, which can be easily converted into a vector $\bm{\Lambda^*}$ of VaR estimates. 
So, inserting $\bm{L^*}$ in equation (\ref{FFTVaR}), we are quickly and efficiently provided with the full relation between the VaR estimates 
and the corresponding appropriate significance levels. Analogously, the FFT computation of $G_{\nu}(L^*,\nu+1)$ provides us with the $\bm{E^*}$ 
spectrum over a whole range of significance levels. Thus, equations (\ref{FFTVaR}) and (\ref{FFTES}) provide a global information about the VaR 
and ES distributions over a wide range of $\pstar$ values. These results lead to a very intuitive and easy to read graphical representation that
we shall discuss in the final section.

Let us now mention that the real part of the $G_{\nu}$ function can be put into a different form. In fact, we can define the function
\begin{widetext}
\begin{eqnarray} \label{I}
  I_{\nu}(L^*,\theta) \doteq \mathrm{Re} \ G_{\nu} (L^*,\theta) &=& \ue^{-\theta L^*} \int_0^{+\infty} \frac{\ud \omega}{\theta^2 + \omega^2} 
  \Big \{ \cos (\omega L^*) \big [ \theta \ \mathrm{Re} f (\omega + \ui \nu) - \omega \ \mathrm{Im} f (\omega + \ui \nu) \big ] \Big \} \nonumber \\ 
  &-& \ue^{-\theta L^*} \int_0^{+\infty} \frac{\ud \omega}{\theta^2 + \omega^2} \Big \{ \sin (\omega L^*) \big [ \omega \ \mathrm{Re} f (\omega + \ui \nu) 
  + \theta \ \mathrm{Im} f (\omega + \ui \nu) \big ] \Big \}
\end{eqnarray}
\end{widetext}
and we can consequently rewrite the equations for risk measures as
\begin{equation} \label{trapz}
  \pstar = \frac{I_{\nu}(L^*,\nu)}{\pi} \ ,  \ \ \es = 1 - e^{\mu \Delta t} \frac{I_{\nu}(L^*,\nu+1)}{I_{\nu}(L^*,\nu)}.
\end{equation}
Now, the $I_{\nu}$ function in ($\ref{I}$), being the sum of sine and cosine transforms, is perfectly suited to numerical evaluation by means of 
trapezoidal integration algorithms, and this partially prevents the risk estimates in (\ref{trapz}) from being affected by the FFT approximations,
which, in turn, can be made negligible only by setting a large $\omega$ grid. The equations in (\ref{trapz}) may prove to be very helpful to the 
evaluation of risk measures at a few specific values of the significance level $\pstar$ with a very high precision. It is also worth stressing again 
that, remarkably, the only input those equations require is the ECF of the financial dynamics under study. 
As we shall discuss in the next section, this fact makes it possible to introduce and successfully employ a number of models in the framework of risk management. 


\section{Models}

One of the crucial points in risk analysis is the evaluation of risk measures over different time lags. So, generally, a projection over 
the desired time horizons of the PDFs employed to model financial returns has to be performed. The most interesting case, being the one 
required by regulators \cite{Basel:2006}, is to project from one to ten trading days. Equations (\ref{FFTVaR}), (\ref{FFTES}) and 
(\ref{trapz}) provide a very natural framework where different time scaling behaviors can be compared. 
In this section we discuss those arising when considering two class of models: the first one of purely additive processes governed by the CLT
and a second class of SVMs.

As it has already been stressed, the main ingredient required by our approach is the knowledge of the closed-form ECF associated with the model in use.
As a representative member of the class of additive process, we consider a very simple one given by an arithmetic motion whose driving noise is described by a TLD 
\cite{Mantegna_Stanley:1994}. The most appealing feature of the TLD is the ability to reproduce some of the 
stylized facts commonly observed in financial markets, such as the asymmetry and the excess of kurtosis \cite{Bouchaud_Potters:2003}. 
As it is well known, the time scaling is governed by the CLT and this causes high order normalized cumulants to decrease monotonically with time. The scaling behavior is
governed by a power law whose characteristic exponent is specific for the order of the cumulant, e.g. the skewness scales as $t^{-0.5}$ while the kurtosis scales 
according to the $t^{-1}$ law.  
On the other hand, SVMs are naturally provided with a different, exponentially damped time scalings, not necessarily monotonic, 
which could be able to better capture real market data time scalings. In this article we want to test the ability of these models to
capture the projection over horizon of risk measures in comparison with historical estimates and the standard Log-Normal dynamics. 
Since the most commonly used SVMs are well characterized in terms of the CF, formulae (\ref{FFTVaR}) and (\ref{FFTES}) allow for a proper comparison and 
the extension of the risk analysis to the context of these models does represent the second main contribution of our work. 
In the following we focus on the Heston model \cite{Heston:1993}, since it represents a benchmark model in the option pricing framework. 
However, our approach could also be easily applied to other SVMs \cite{Stein_Stein:1991,Schobel_Zhu:1999,Bormetti_etal:2008,Bormetti_Cazzola_Delpini:2009} 
and extended models dealing, for example, with jump diffusion \cite{Lipton_Sepp:2008}. 

\subsection{Truncated L\'evy Distributions}

The CF of a TLD can be expressed as $f(\phi) = \exp (H(\phi))$, where the cumulant generating function, or Hamiltonian, $H$ is given by \cite{Koponen:1995,Kleinert:2002,Kleinert:2009}
\begin{widetext}
\begin{equation}\label{eq:Ham}
  H(\phi) = - \frac{\Sigma^2}{2} \frac{\lambda^{2-\gamma}}{\gamma(1-\gamma)} \Big \{ (1+\beta) \exp \big [ \gamma \log (\lambda + \ui \phi) \big ] 
  + (1-\beta) \exp \big [ \gamma \log (\lambda - \ui \phi) \big ] - 2 \lambda^{\gamma} \Big \}
\end{equation}
\end{widetext}
and this gives rise to the following asymptotic behavior for the PDF $p_{TL}$:
\begin{equation}\label{eq:asymp}
  p_{TL}(x) \overset{|x| \to + \infty}{\longrightarrow} C_{\Sigma,\gamma,\lambda,\beta} \frac{\ue^{-\lambda |x|}}{|x|^{1+\gamma}} [1 + \beta \ \mathrm{sign} (x)].
\end{equation}
$C_{\Sigma,\gamma,\lambda,\beta}$ is a constant depending on the four free parameters which define the distribution. 
As it is clear from ($\ref{eq:asymp}$), $\beta$ determines the level of asymmetry of the PDF, $\gamma \in (0,2]$ is the tail exponent 
($\gamma = 2$ reducing to the Normal case), while $\lambda > 0$ is the decay factor; $\Sigma >0$ defines the level of the second moment of the distribution. 
The values of such parameters single out a particular TLD univocally. However, in our case, a corrective positional term has to be added to the Hamiltonian 
function ($\ref{eq:Ham}$). This is because  we work with centered log-returns empirical distribution and so we need to correctly center 
the TLD on the real data. Thus we consider the modified Hamiltonian $H^{'}(\phi)$
\begin{equation}
	H^{'}(\phi) = H(\phi) - \ui \left( k_1 + \frac{\Sigma^2}{2} \right) \phi
\end{equation}
where $k_1=-\ui \left.\frac{\ud H}{\ud \phi}\right|_{\phi=0}$. 
The singularity of the model relevant for the strip of regularity is readily found by solving the equation $\omega+\ui\nu_+=\ui \lambda$.\\
The presence of an exponential damping in ($\ref{eq:asymp}$) guarantees the finiteness of the variance, which can be shown to be equal to $\Sigma^2$, 
and of all the higher order moments. The CLT applies and this has immediate consequences on the time scaling of many relevant quantities. 
As a matter of fact, upon the addition of $N$ independent identically distributed (\emph{i.i.d.}) TLD variables the generic $n$-th order cumulant $k_n$ scales linearly with $N$. 
Then, for the skewness $\zeta$ and kurtosis $\kappa$ we have
\begin{equation}
  \zeta(N) = \frac{k_3(N)}{k_2^{3/2}(N)} \sim N^{-1/2} \ , \ \ \ \kappa(N) = \frac{k_4(N)}{k_2^2(N)} \sim N^{-1}.
\end{equation}
As it will be discussed in the next paragraph, the Heston stochastic volatility model leads to a much richer time evolution of the cumulants.

\subsection{Heston dynamics}

The Heston model is defined by the two following coupled stochastic differential equations (SDEs)
\begin{eqnarray} 
  \ud S(t) &=& \mu S(t) \ud t + \sqrt{v(t)} S(t) \ud W_1(t) \\ \label{HestonS}
  \ud v(t) &=& \alpha (\sigma^2 - v(t)) \ud t + k \sqrt{v(t)} \ud W_2(t) \label{Hestonv}
\end{eqnarray} 
under the initial conditions $S(0)=S_0$, $v(0)=\sigma^2$, and for strictly positive $\alpha$ and $k$.
The noise $W_2(t)$ is defined in terms of the standard Brownian increments $\ud W_1(t)$ and $\ud W(t)$ through the usual relation 
$\ud W_2(t) = \rho \ud W_1(t) + \sqrt{1-\rho^2} \ud W(t)$ and $\rho\in[-1;1]$. 
In the following, we shall work with centered log-returns $X$, whose SDE can be derived from the previous ones:
\begin{equation}
  \ud X(t) = -\frac{v(t)}{2} \ud t + \sqrt{v(t)} \ud W_1(t)
\end{equation}
with $X_0 = 0$. 
By means of standard techniques \cite{Heston:1993,Dragulescu_Yakovenko:2002,Lord:2006} the cumulant generating function of the model can be shown to be
\begin{widetext}
\begin{equation} \label{HestonH}
  H(\phi) = \frac{\alpha \sigma^2}{k^2} \Big [ \big (\xi(\phi) - \eta(\phi) \big) t - 2 \ln \big (1 - g(\phi) \ue^{-\eta(\phi) t } \big ) 
  + 2 \ln \big (1 - g(\phi) \big ) \Big ] + \frac{\sigma^2 \big (\xi(\phi) - \eta(\phi) \big )}{k^2} \frac{1 - \ue^{-\eta(\phi) t}}{1 - g(\phi) \ue^{-\eta(\phi) t}}
\end{equation}
\end{widetext}
with
\begin{eqnarray*}
  \xi (\phi) &=& \alpha - \ui \rho k \phi \\
  \eta (\phi) &=& \sqrt{\xi^2(\phi) + k^2 \phi (\ui + \phi)} \\
  g(\phi) &=& \frac{\xi(\phi) - \eta(\phi)}{\xi(\phi) + \eta(\phi)}.
\end{eqnarray*}
From ($\ref{HestonH}$) the cumulants of the Heston model PDF can be derived explicitly, and their analytical expressions are reported in Appendix \ref{app:Heston}.
The identification of the strip $(0,\nu_+)$ is more tricky for the Heston case. The relevant singular points solve the equations $\eta(\ui\nu_+^a)=0$ and 
$1-g(\ui\nu_+^b)\ue^{-\eta(\ui\nu_+^b)t}=0$. Restricting to the case $\rho\in(-1,1)$, from the former equation we have 
$\nu_+^a=\left[2\alpha\rho-k+\sqrt{(2\alpha\rho-k)^2+4\alpha^2(1-\rho^2)}\right]/\left[2k(1-\rho^2)\right]$; the latter can not be solved explicitly 
but, once a set of parameters values has been fixed, we can numerically check if there is any positive $\nu_+^b<\nu_+^a$. 
However, for typical values of $\alpha\sim 10^2$ and $k\sim 10$, and for $\nu\in(0,1]$ we obtain positive $\eta$ and $\xi$ of order $10^2$;
so, for each $t>0$, a candidate $\nu_+^b$ has to be strictly greater than one. Thus, if we fix $\nu=1$ when integrating equations~(\ref{FFTVaR}) and (\ref{FFTES}),
we have only to check that $\nu_+=\nu_+^a>1$.


\section{Data analysis and numerical results}

As already remarked, our aim is to capture the empirical scaling of the returns and to exploit it in order to end up with reliable risk 
estimates projections over the time horizons of interest. Both in the TLD and in the Heston model case, we perform this task by focusing on the calibration
on the first four cumulants time scalings.

\subsection{Data Sets and Calibration}

TLDs are fitted on empirical distributions by means of a simple step-by-step procedure according to which the free parameters in ($\ref{eq:Ham}$), $\Sigma$,
$\gamma$, $\beta$ and $\lambda$, are evaluated one at a time. 
First, extending the approaches developed in \cite{Mantegna_Stanley:1994,Matacz:2000} to the case of asymmetric TLDs, 
the tail exponent $\gamma$ is fitted exploiting the time scaling of the empirical zero return probability. 
Actually, the quantity $p_L(X=0)$ of a L\'evy distribution (a good approximation for the central region of a TLD, since the exponential damping mainly affects 
the tail regions) is used. Such a point can be shown to scale linearly with time $t$ on a log-log scale
\begin{equation}
  \log p_L^{(t)}(0) = \log \frac{f(\gamma,\beta)}{c \pi \gamma} - \frac{1}{\gamma} \log t
\end{equation}
where $c$ is a constant, while $f$ is a function of the $\gamma$ and $\beta$ parameters. 
Clearly, from the slope of this linear relation the value of $\gamma$ can be estimated. 
Moreover, we consider the following relations \cite{Kleinert:2002} valid for the variance, skewness and kurtosis of a TLD over an horizon $t$
\begin{eqnarray} \label{fit}
  k_2(t) &=& \Sigma^2 t \nonumber \\
  \kappa(t) &=& (2 - \gamma) (3 - \gamma) / (\lambda^2 \Sigma^2 t) \nonumber \\
  \zeta(t) &=& \beta (2 - \gamma) / (\lambda \sqrt{\Sigma^2 t}).
\end{eqnarray}
$\Sigma$ can be readily estimated from the first relation, while $\lambda$ and $\beta$ can be obtained from the remaining ones. 
Both last two relations can be rearranged into linear ones on a log - log scale. Thus, the parameters of interest can all be estimated 
via Marquardt-Levenberg linear fits.

The Heston model calibration is performed differently. 
By imposing the initial condition $v_0 = \sigma^2$, we assumed the model to be in its stationary volatility state. The empirical mean $\mu$ is estimated from the 
linear returns directly. Moreover, it is clear from the analytical expression reported in Appendix \ref{app:Heston} that $\sigma$ can be fitted on the time 
scaling of the first cumulant. 
This leaves us with three more parameters to be estimated, i.e. $\alpha$, $\rho$ and $k$. We obtain the optimal values by solving the following minimization problem numerically
\begin{equation*}
  \alpha^*,\rho^*,k^* = \underset{\rho \in (-1,1); \, \alpha, k > 0}{\mathrm{argmin}} \sum_{j=1}^{10} \sum_{i=2}^4 
  \left [ \frac{k_i^E (j\Delta t) - k_i (j\Delta t)}{\epsilon_{k_i} (j\Delta t)} \right ]^2.
\end{equation*}
In the previous equation the first sum runs over trading days ($\Delta t = 3.98\times 10^{-3}$ years), while the second one runs over the cumulants: 
$k_i^E$ represents the $i-$th empirically estimated cumulant, with statistical uncertainty $\epsilon_{k_i}$, 
whereas $k_i$ represents the $i-$th analytical cumulant.

The calibration has been performed on three different data sets, made of 5000 daily returns each, from the German DAX 30 Index (from November 14th 1988 to September 9th 2008), 
the French CAC 40 Index (from November 10th 1988 to September 9th 2008) and the Dow Jones EURO STOXX 50 (SX5E) Index (from March 10th 1989 to September 9th 2008). 
In Table~\ref{tab:Params} all of the relevant parameter estimates, both for the TLD and the Heston model, are reported. 
\begin{table*}
  \caption{\label{tab:Params} Values of $\mu$ (left) and of the TLD (center) and Heston (right) models parameters as estimated from the data sets.}
  \begin{ruledtabular}
  \begin{tabular}{l c | cccc | cccc}
  & $\mu (\times 10^{-2})$ & $\Sigma^2$ ($\%$) & $\gamma$ & $\lambda$ & $\beta$ & $\sigma^2$ ($\%$) & $\alpha$ & $k$ & $\rho$ \\
  \hline
  DAX  & 11.02 & 4.64 & 1.77 & 10.74 & -0.38 & 4.71 & 86 & 4.67 & -0.17 \\
  CAC  & 7.47 & 4.11 & 1.84 & 11.78 & -0.21 & 4.21 & 330 & 8.08 & -0.06 \\
  SX5E & 8.73 & 3.55 & 1.78 & 13.60 & -0.33 & 3.88 & 287 & 8.82 & -0.12 \\
  \end{tabular}
  \end{ruledtabular}
\end{table*}
\begin{figure*}
  \includegraphics[height=5 cm,width=8 cm]{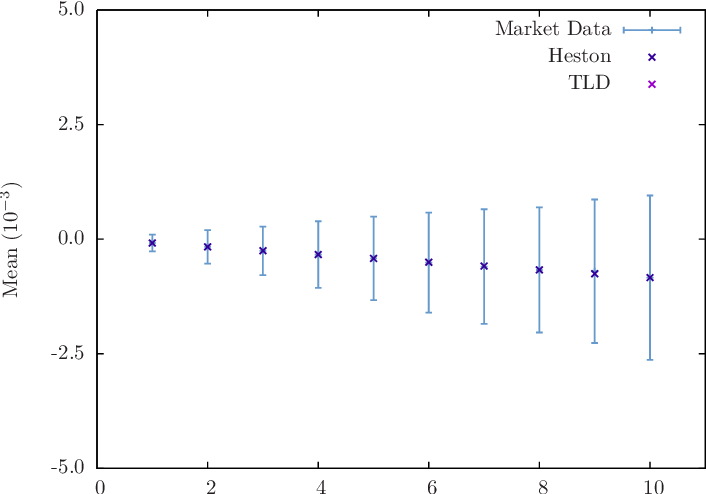}
  \hspace{0.05 in}
  \includegraphics[height=5 cm,width=8 cm]{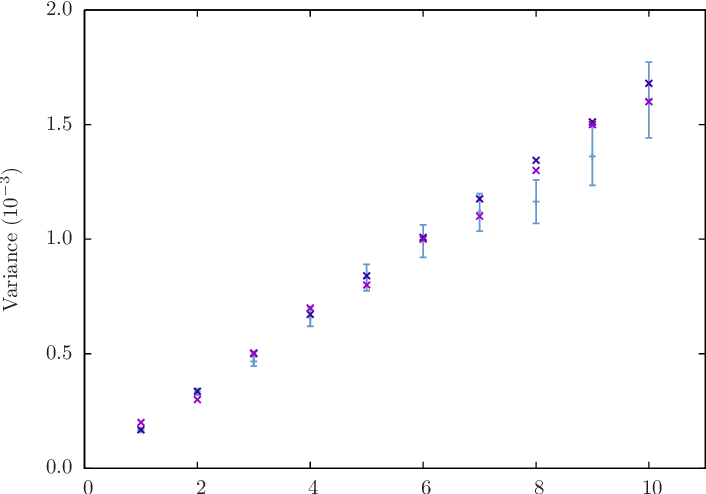}\\
  \vspace{0.1 in}
  \includegraphics[height=5 cm,width=8 cm]{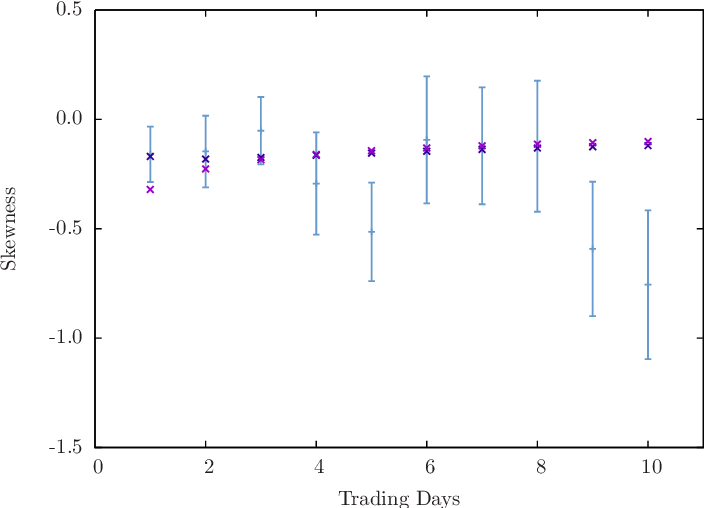} 
  \hspace{0.05 in}
  \includegraphics[height=5 cm,width=8 cm]{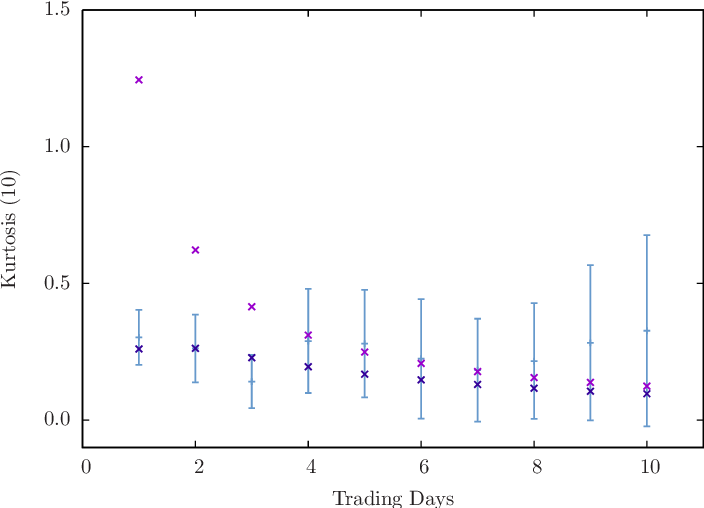}
    \caption{\label{fig:CACScaling} Comparison between the empirical, TLD and Heston time scaling of the mean, variance, 
    skewness and kurtosis for the CAC 40 Index time series over a 10 days horizon. 
    }
\end{figure*}
\begin{figure*}
  \includegraphics[height=6 cm,width=8 cm]{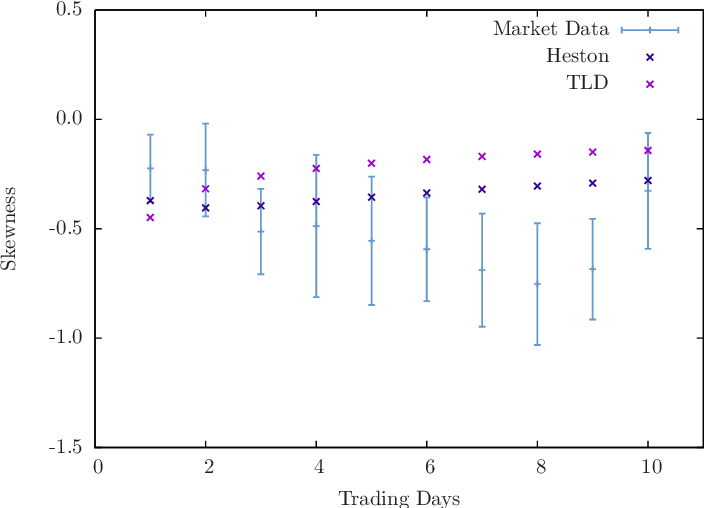}
  \hspace{0.1 in} 
  \includegraphics[height=6 cm,width=8 cm]{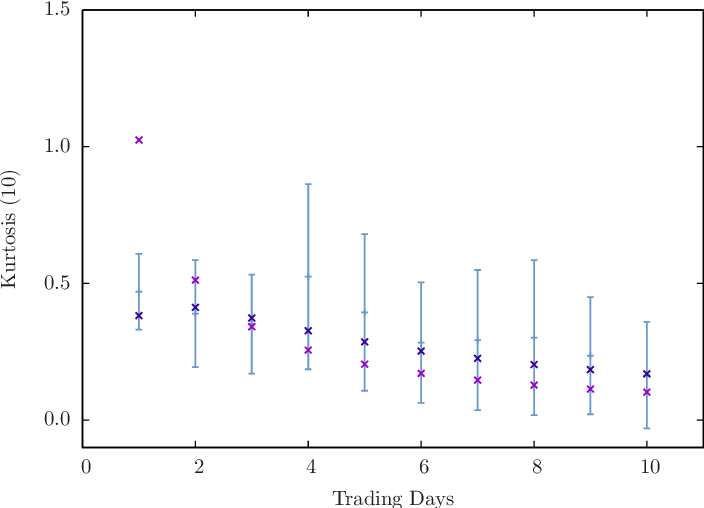} 
  \caption{\label{fig:SXEScaling} Comparison between the empirical, TLD and Heston time scaling of the skewness and kurtosis for the SX5E Index over a 10 days horizon.}
\end{figure*}
In Figure~\ref{fig:CACScaling} the different time scalings obtained from such parameters are compared with the empirical ones. 
In the two upper figures the time scaling of the mean and variance is considered. 
An excellent agreement is observed for both the TLD and the Heston model. 
In the lower figures, even though the calibration was performed over the cumulants, we report the skewness and kurtosis 
time scaling because of their major relevance in risk analysis. Due to error propagation, the error bars look much more irregular in these cases. 
As it can be seen, the Heston model better describes the empirical scaling than the TLD model, especially for the kurtosis data. For the skewness scaling, Heston slightly 
outperforms the TLD approach, as confirmed in Figure~\ref{fig:SXEScaling} for the SX5E Index case. We shall see in the next section how this behaviour is translated in terms of
the risk estimates.

\subsection{Risk Estimates}

In this section we detail and discuss all of the risk estimates we obtained. We present the results based on the TLD and Heston models under the generalized Fourier transform 
approach. As already discussed, this essentially amounts to the use of formulae (\ref{G}), (\ref{FFTVaR}) and (\ref{FFTES}). As benchmark models we consider the 
Log-Normal dynamics and the historical approach. The former is meant to provide a comparison with the results obtained under the standard normality assumption for the returns;
the latter, instead, provides some insight into the actual risk levels of an asset. 
Historical estimates are obtained with standard methodologies (see \cite{Bormetti_etal:2007} for example).

We also employ the calibration procedure described in the above section in a bootstrap framework in order to provide the risk estimates with $68\%$ CL intervals. 
For each data set we generate $M_B=1000$ synthetic copies of the original time series and this is done 
by means of a GARCH(1,1) model to preserve the correlation structure of the volatility. Such a model is defined by the following couple of equations
\begin{eqnarray} \label{eq:GARCH}
  Y_t &=& C + \sigma_t z_t \nonumber \\
  \sigma^2_t &=& K + G \sigma^2_{t-1} + A \sigma^2_{t-1} z^2_{t-1},
\end{eqnarray}
where $Y_t$ is the log-return at time $t$, $\sigma_t$ describes its volatility and the $z_t$'s, often referred to as innovations, correspond to a Gaussian white noise;
$C$, $K$, $G$ and $A$ are constant quantities. 
The model calibration can be succesfully performed with the help of R software (\verb|www.r-project.org|). 
Being \emph{i.i.d.} Gaussian variables, the bootstrap technique can be applied to the innovations in order to generate replicas of the original 
time series preserving the volatility clustering. In our analysis, the TLD and Heston model calibrations have been carried out on each bootstrap copy. 
This provides us with a different set of parameter values for each copy that can be plugged into the ECF and consequently into equations (\ref{FFTVaR}) and (\ref{FFTES}) 
to obtain copy-dependent risk estimates $\theta^*_j$ for $j=1, \ldots, M_B$, where $\theta$ can represent either VaR or ES. 
Then, bootstrap confidence levels are defined as $[\theta^*_\alpha;\theta^*_{1-\alpha}]$, with the boundaries of the interval satisfying the following relation:
$\mathrm{Prob}(\theta^* \leq \theta^*_{\alpha, 1-\alpha}) = \alpha, 1 - \alpha$. Thus, a $68\%$ CL interval requires $\alpha=16\%$.
The bootstrap technique allows us to draw statistically robust conclusions. 
\begin{figure*}
  \includegraphics[height=6 cm,width=8 cm]{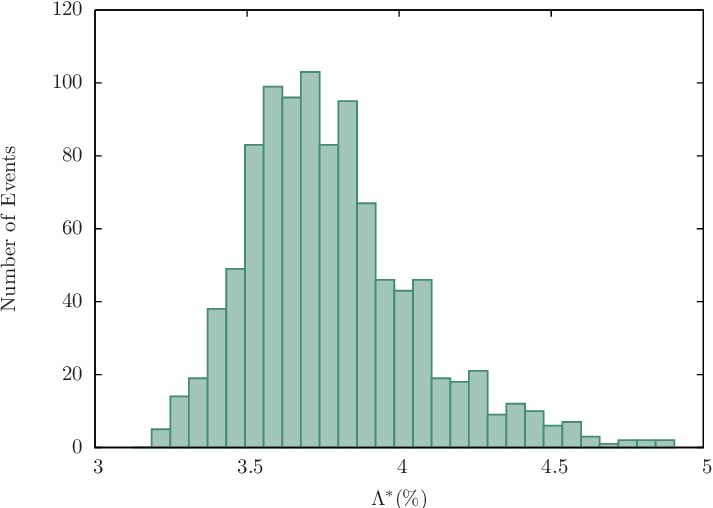}
  \hspace{0.1 in} 
  \includegraphics[height=6 cm, width=8 cm]{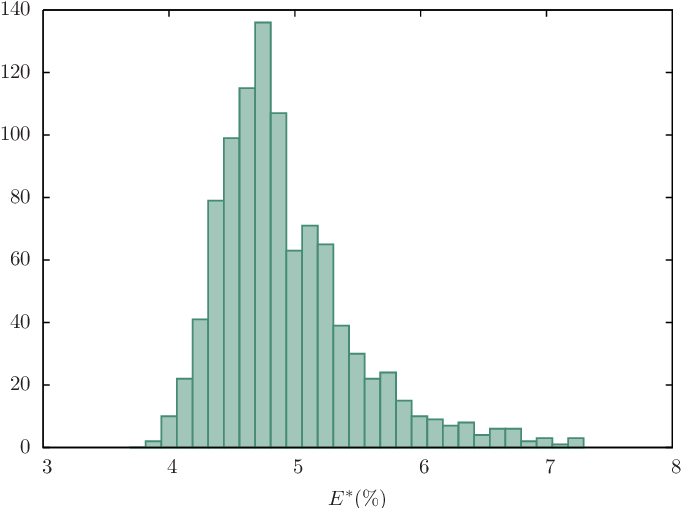} 
  \caption{\label{fig:Hist} VaR and ES bootstrap histograms for the Heston model; both plots refer to $\pstar=1\%$ over a 1 day horizon for the CAC 40 Index.}
\end{figure*}
In Figure~\ref{fig:Hist} we report two example bootstrap histograms of $\VaR_H$ and $\es_H$ for the CAC 40 Index; from their analysis   
we can identify the extremes of the confidence intervals at the desired CL.

As already mentioned, one of the main motivations to our approach is the need for a natural projection over horizon framework. 
As required by regulators, we focus on risk measures evaluation at 1 day and 10 days horizons. 
Besides this, VaR and ES are evaluated at $\pstar = 1\%$, this also being requested by regulators, and at $\pstar = 5\%$.
\begin{table}
\caption{\label{tab:1D1CL} Historical, Normal, TLD and Heston VaR and ES estimates at $\pstar=1\%$ over a 1 day horizon.}
\begin{ruledtabular}
\begin{tabular}{lcccc}
& $\VaR_{\mathrm{Hist}}$ $(\%)$  & $\VaR_N$ ($\%$) & $\VaR_T$ ($\%$) & $\VaR_H$ ($\%$) \\
\hline
DAX & $3.94_{-0.17}^{+0.18}$ & $3.10_{-0.10}^{+0.09}$ & $3.36_{-0.28}^{+0.28}$ & $3.69_{-0.39}^{+0.40}$ \\
CAC & $3.54_{-0.15}^{+0.15}$ & $2.95_{-0.07}^{+0.07}$ & $3.01_{-0.20}^{+0.20}$ & $3.53_{-0.26}^{+0.26}$ \\
SX5E& $3.62_{-0.16}^{+0.15}$ & $2.82_{-0.09}^{+0.09}$ & $3.16_{-0.26}^{+0.26}$ & $3.61_{-0.33}^{+0.36}$\\
\hline \hline
& $\es_{\mathrm{Hist}}$ $(\%)$  & $\es_N$ ($\%$) & $\es_T$ ($\%$) & $\es_H$ ($\%$) \\
\hline
DAX & $5.25_{-0.40}^{+0.42}$ & $3.60_{-0.12}^{+0.10}$ & $4.78_{-0.59}^{+0.57}$ & $4.52_{-0.87}^{+0.90}$ \\
CAC & $4.55_{-0.29}^{+0.27}$ & $3.38_{-0.07}^{+0.07}$ & $4.03_{-0.40}^{+0.43}$ & $4.44_{-0.48}^{+0.46}$ \\
SX5E& $4.71_{-0.39}^{+0.35}$ & $3.25_{-0.09}^{+0.11}$ & $4.67_{-0.50}^{+0.54}$ & $4.63_{-0.73}^{+0.84}$\\
\end{tabular}
\end{ruledtabular}
\end{table}
\begin{table}
\caption{\label{tab:10D1CL} Historical, Normal, TLD and Heston VaR and ES estimates at $\pstar=1\%$ over a 10 days horizon.}
\begin{ruledtabular}
\begin{tabular}{lcccc}
& $\VaR_{\mathrm{Hist}}$ $(\%)$  & $\VaR_N$ ($\%$) & $\VaR_T$ ($\%$) & $\VaR_H$ ($\%$) \\
\hline
DAX & $13.01_{-1.52}^{+1.56}$ & $9.27_{-0.34}^{+0.36}$ & $9.38_{-0.63}^{+0.57}$ & $11.71_{-1.07}^{+1.07}$ \\
CAC & $11.98_{-1.42}^{+1.30}$ & $8.89_{-0.29}^{+0.27}$ & $8.72_{-0.43}^{+0.44}$ & $9.80_{-0.70}^{+0.71}$ \\
SX5E& $9.85_{-1.45}^{+1.44}$ & $8.50_{-0.34}^{+0.34}$ & $8.76_{-0.62}^{+0.65}$ & $9.95_{-0.95}^{+1.03}$\\
\hline \hline
& $\es_{\mathrm{Hist}}$ $(\%)$  & $\es_N$ ($\%$) & $\es_T$ ($\%$) & $\es_H$ ($\%$) \\
\hline
DAX & $16.39_{-2.21}^{+2.16}$ & $10.58_{-0.38}^{+0.38}$ & $11.57_{-1.07}^{+1.03}$ & $14.81_{-1.65}^{+1.55}$ \\
CAC & $15.12_{-1.76}^{+1.81}$ & $10.13_{-0.30}^{+0.29}$ & $10.48_{-0.62}^{+0.65}$ & $11.83_{-1.06}^{+1.05}$ \\
SX5E& $11.16_{-2.30}^{+2.12}$ & $9.69_{-0.37}^{+0.37}$ & $11.07_{-1.01}^{+1.09}$ & $12.28_{-1.47}^{+1.61}$\\
\end{tabular}
\end{ruledtabular}
\end{table}
\begin{table}
\caption{\label{tab:1D5CL} Historical, Normal, TLD and Heston VaR and ES estimates at $\pstar=5\%$ over a 1 day horizon.}
\begin{ruledtabular}
\begin{tabular}{lcccc}
& $\VaR_{\mathrm{Hist}}$ $(\%)$  & $\VaR_N$ ($\%$) & $\VaR_T$ ($\%$) & $\VaR_H$ ($\%$) \\
\hline
DAX & $2.13_{-0.06}^{+0.06}$ & $2.19_{-0.07}^{+0.07}$ & $2.01_{-0.08}^{+0.08}$ & $2.28_{-0.10}^{+0.13}$ \\
CAC & $2.08_{-0.05}^{+0.05}$ & $2.09_{-0.05}^{+0.05}$ & $1.93_{-0.06}^{+0.06}$ & $2.08_{-0.07}^{+0.08}$ \\
SX5E& $1.91_{-0.05}^{+0.05}$ & $1.99_{-0.07}^{+0.07}$ & $1.78_{-0.06}^{+0.07}$ & $2.01_{-0.11}^{+0.11}$\\
\hline \hline
& $\es_{\mathrm{Hist}}$ $(\%)$  & $\es_N$ ($\%$) & $\es_T$ ($\%$) & $\es_H$ ($\%$) \\
\hline
DAX & $3.26_{-0.12}^{+0.12}$ & $2.75_{-0.09}^{+0.08}$ & $2.95_{-0.21}^{+0.20}$ & $3.17_{-0.24}^{+0.24}$ \\
CAC & $3.00_{-0.10}^{+0.10}$ & $2.63_{-0.07}^{+0.07}$ & $2.68_{-0.14}^{+0.14}$ & $3.00_{-0.17}^{+0.17}$ \\
SX5E& $2.94_{-0.11}^{+0.10}$ & $2.51_{-0.09}^{+0.09}$ & $2.71_{-0.19}^{+0.20}$ & $3.01_{-0.21}^{+0.23}$\\
\end{tabular}
\end{ruledtabular}
\end{table}
\begin{table}
\caption{\label{tab:10D5CL} Historical, Normal, TLD and Heston VaR and ES estimates at $\pstar=5\%$ over a 10 days horizon.}
\begin{ruledtabular}
\begin{tabular}{lcccc}
& $\VaR_{\mathrm{Hist}}$ $(\%)$  & $\VaR_N$ ($\%$) & $\VaR_T$ ($\%$) & $\VaR_H$ ($\%$) \\
\hline
DAX & $6.53_{-0.51}^{+0.52}$ & $6.55_{-0.30}^{+0.29}$ & $6.09_{-0.31}^{+0.30}$ & $6.74_{-0.36}^{+0.36}$ \\
CAC & $6.04_{-0.46}^{+0.47}$ & $6.31_{-0.25}^{+0.23}$ & $5.87_{-0.26}^{+0.24}$ & $6.36_{-0.31}^{+0.30}$ \\
SX5E& $5.64_{-0.46}^{+0.45}$ & $6.01_{-0.28}^{+0.28}$ & $5.53_{-0.29}^{+0.29}$ & $6.12_{-0.36}^{+0.36}$\\
\hline \hline
& $\es_{\mathrm{Hist}}$ $(\%)$  & $\es_N$ ($\%$) & $\es_T$ ($\%$) & $\es_H$ ($\%$) \\
\hline
DAX & $9.99_{-0.91}^{+0.87}$ & $8.22_{-0.32}^{+0.33}$ & $8.18_{-0.52}^{+0.52}$ & $9.73_{-0.78}^{+0.80}$ \\
CAC & $9.06_{-0.77}^{+0.76}$ & $7.89_{-0.27}^{+0.26}$ & $7.66_{-0.38}^{+0.35}$ & $8.49_{-0.54}^{+0.55}$ \\
SX5E& $7.70_{-0.86}^{+0.84}$ & $7.52_{-0.32}^{+0.32}$ & $7.60_{-0.50}^{+0.52}$ & $8.49_{-0.71}^{+0.75}$\\
\end{tabular}
\end{ruledtabular}
\end{table}
\begin{figure*}
  \includegraphics[height=6 cm,width=8 cm]{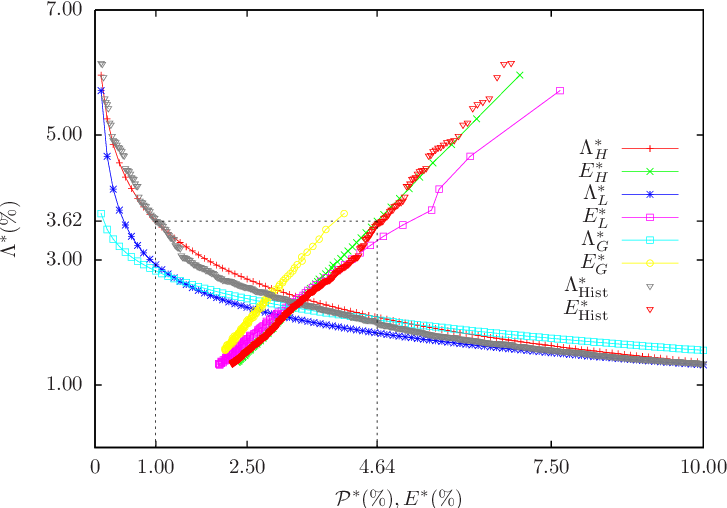}
  \hspace{0.1 in} 
  \includegraphics[height=6 cm, width=8 cm]{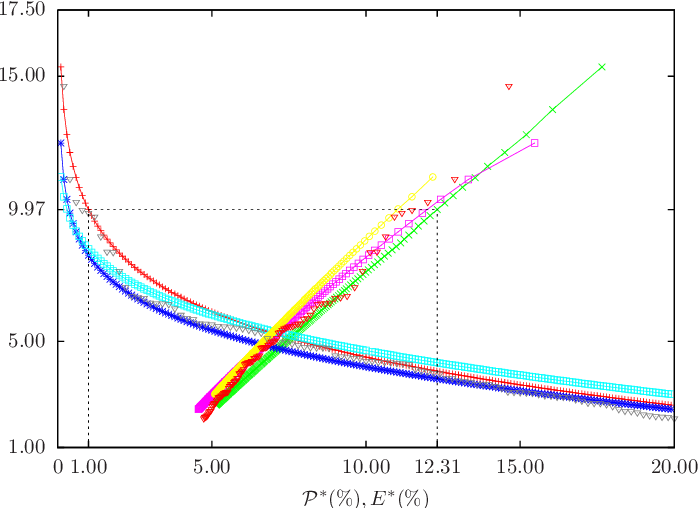} 
  \caption{\label{fig:SX5E} Plot of $\VaR$ \emph{vs} $\pstar$ and $\es$ over 1 day (left) and 10 days (right) horizons for the SX5E Index.} 
\end{figure*}
In Table~\ref{tab:1D1CL} and Table~\ref{tab:10D1CL} the risk estimates for $\pstar = 1\%$ at 1 day and 10 days horizons are detailed; 
the ``Hist'' subscript refers to historical estimates, while $N$ refers to the Normal ones, $T$ and $H$ to the TLD and Heston models, respectively. 
A few comments need now to be made. First of all, the best agreement with historical estimates is found for $\VaR_H$ and $\es_H$, and 
it is remarkable for the $\pstar=1\%$, 1 day VaR. 
The agreement slightly worsens for the TLD, while the Normal estimates widely underestimate $\VaR_{\mathrm{Hist}}$ and $\es_{\mathrm{Hist}}$. 
These results confirm the ability of both the Heston and TLD models to better describe tail events of the empirical distributions than the 
Log-Normal model. When considering a 10 days horizon the errors get much larger, since the empirical risk estimates are evaluated on the basis of 500 returns only. 
The Heston and historical values are in best statistical agreement, while TLD and Normal estimates definitely worsen, and
they get much closer as a consequence of the CLT, as expected. These results suggest that the projection over time horizon associated with the dynamics 
of the Heston model can provide a better description of the risk level for low $\pstar$ than the one induced by the CLT.  
As far as a higher $\pstar$ level is concerned, we consider $\pstar=5\%$ (see Table~\ref{tab:1D5CL} and Table~\ref{tab:10D5CL}).
One can see that, both for the daily and 10 days horizon, the difference between the different VaR estimates reduces. 
For the ES, which by definition exibits a higher sensitivity to the tail behaviour than VaR, 
we again find the best performance for the Heston model. 
In particular, when switching to the 10 days horizon, the Heston model definitely leads to the best overall agreement with historical estimates.
Finally, we present a graphical representation, very effective for practical applications. 
In Figure~\ref{fig:SX5E}, left panel, we plot $\VaR$ against the significance level over a daily horizon; the curves are obtained via 
an adaptive trapezoidal integration of the $I_\nu(L^*,\theta)$ function to increase the numerical accuracy w.r.t. FFT based approaches. 
We consider a grid of one hundred equally spaced $\pstar$ values ranging from $0.1\%$ to $10\%$ and $\nu=1$. Triangles represent historical estimates and the 
matching with the Heston curve is quite evident. We explicitly draw a vertical dotted line for $\pstar=1\%$. 
The crossing point with the Heston curve identifies the VaR estimate for the specified significance level. 
If we translate the estimate over the ES curve we obtain a new crossing point, whose projection over the horizontal axis returns the desired ES value. 
In the right panel of the same figure the analysis is performed for a 10 days horizon. 
This graphical approach shows how straightforward it is to obtain the risk estimates and to compare performances provided by different models.

\section{Conclusions}

In this paper we have shown how the extension of the CF to the complex domain can be successfully employed to derive compact formulae 
describing VaR and ES, the two market risk measures most financial industries ordinarily use. The same technique has already been widely adopted and tested
in the context of option pricing under stochastic volatility. The analogy is not so surprising when noticing that the integral equation linking $L^*$ and $\pstar$,
see equation~(\ref{VaR2}), clearly resembles the relation between the strike and the price of an option with a digital payoff. A similar argument applies to the
ES, even though in this case the payoff function looks a little bit more complicated, 
i.e. $\left[1-\Theta(x+L^*)\right]\left(\ue^{x+\mu\Delta t}-1\right)$, where $\Theta$ is the Heaviside step function, see equation~(\ref{eq:ES}).
Exploiting these analogies, we have obtained new integral representation of risk in terms of the function $G_\nu(L^*,\theta)$, particularly suited 
to efficient numerical integration using FFT algorithms. Based on the function $I_\nu(L^*,\theta)$ and on adaptive trapezoidal algorithms, 
we have also suggested an alternative approach to perform an accurate integration of our formulae. 
Our focus was on two types of log-returns stochastic dynamics. The first one is a simple arithmetic motion whose random increments correspond
to a Truncated L\'evy noise, while the second is the Heston stochastic volatility dynamics. Both models are analytically well-defined in terms
of the CF; moreover, the CLT applies to the former predicting a power-law scaling with time of high order normalized cumulants. In the Heston case, instead,
their time evolution obeys an exponentially damped scaling. Since the risk measure projection over time horizons is one of the points addressed by regulators,
we have compared the performances provided by these conceptually quite different models.
We have tested our analytical formulae on a data set of financial indexes, the German DAX 30 Index, the French CAC 40 and the European Dow Jones EURO STOXX 50. 
The results we obtained both for the TLD and Heston cases have shown an excellent agreement with the historical benchmark values, within the bootstrap 68\% CL . 
In particular, at the significance level required by regulators, i.e. $\pstar=1\%$, the Heston model was found to provide the best risk characterization, at least 
for the data we took into account.

Possible perspectives of the present work concern the application of the approach
to the evaluation of risk measures using other SVMs, such as exponential Ornstein-Uhlenbeck
models and their extensions with jump diffusion.

\appendix

\section{Heston model cumulants}\label{app:Heston}
In the following we report the analytical expressions of the Heston model cumulants
\begin{widetext}
  \begin{eqnarray*}
	  k_1&=&-\frac{1}{2}\sigma^2 t,\\
	  k_2&=&\frac{\sigma^2}{8\alpha^3}\big[-k^2\ue^{-2\alpha t}+4k\ue^{-\alpha t}(k-2\alpha\rho)+2\alpha t(4\alpha^2+k^2-4\alpha k\rho)+k(8\alpha\rho-3k)\big],\\
	  k_3&=&\frac{k\sigma^2}{8\alpha^5}\Big\{k^3\ue^{-3\alpha t}-3\alpha k\ue^{-2\alpha t}\left[-k t (k-2\alpha\rho)-2(\alpha-k\rho)\right]
	  -3\ue^{-\alpha t}\left[2\alpha k t (k-2\alpha\rho)^2+3k^3-8\alpha^3\rho-16\alpha k^2\rho\right.\\
	  &&\left.+8\alpha^2k(1+2\rho^2)\right]+3\alpha t\left[-k^3+8\alpha^3\rho+6\alpha k^2\rho-4\alpha^2k(1+2\rho^2)\right]
	  +8k^3-24\alpha^3\rho-42\alpha k^2\rho+6\alpha^2k(3+8\rho^2)\Big\},\\
	  k_4&=&\frac{3k^2 \sigma^2}{64\alpha^7}\Big\{-3k^4\ue^{-4\alpha t}-8k^2\ue^{-3\alpha t}\left[2\alpha k t(k-2\alpha\rho)
	  +4\alpha^2+k^2-6\alpha k\rho\right]-4\ue^{-2\alpha t}\Big[4\alpha^2k^2 t^2 (k-2\alpha\rho)^2\\
	  &&+2\alpha k t\big[k^3-16\alpha^3\rho-12\alpha k^2\rho
	  +4\alpha^2 k(3+4\rho^2)\big]+8\alpha^4-3k^4-32\alpha^3k\rho+8\alpha k^3\rho+16\alpha^2k^2\rho^2\Big]\\
	  &&-8\ue^{-\alpha t}\Big[-2\alpha^2k t^2(k-2\alpha\rho)^3-8\alpha t\big[k^4-7\alpha k^3\rho+4\alpha^4\rho^2
	  -8\alpha^3k\rho(1+\rho^2)+\alpha^2k^2(3+14\rho^2)\big]-9k^4+70\alpha k^3\rho\\
	  &&+32\alpha^3k\rho(4+3\rho^2)-16\alpha^4(1+4\rho^2)-4\alpha^2k^2(9+40\rho^2)\Big]
	  +4\alpha t \big[5k^4-40\alpha k^3\rho-32\alpha^3k\rho(3+2\rho^2)+16\alpha^4(1+4\rho^2)\\
	  &&+24\alpha^2k^2(1+4\rho^2)\big]-73k^4+544\alpha k^3\rho+128\alpha^3k\rho(7+6\rho^2)
	  -32\alpha^4(3+16\rho^2)-64\alpha^2k^2(4+19\rho^2)\Big\}.
  \end{eqnarray*}
\end{widetext}

\end{document}